\documentclass{ws-ijprai}
\usepackage{multirow}

\begin{document}

\markboth{Xueyu Liu,Jie Lin and Chao Wang }{Graph-based multi-Feature fusion method for speech emotion
recognition}

%%%%%%%%%%%%%%%%%%%%% Publisher's area please ignore %%%%%%%%%%%%%%%
%
\catchline{}{}{}{}{}
%
%%%%%%%%%%%%%%%%%%%%%%%%%%%%%%%%%%%%%%%%%%%%%%%%%%%%%%%%%%%%%%%%%%%%

\title{Graph-based multi-Feature fusion method for speech emotion recognition}

\author{Xueyu Liu}
\address{School of Economics and Management,Tongji University, No.1 Zhangwu Road\\
Shanghai,200090,China\,\\
\email{2210458@tongji.edu.cn}}

\author{Jie Lin\footnote{Corresponding author.}}
\address{School of Economics and Management,Tongji University, No.1 Zhangwu Road\\
Shanghai,200090,China,\\
\email{linjie@tongji.edu.cn}
}

\author{Chao Wang}
\address{School of Economics and Management,Tongji University, No.1 Zhangwu Road\\
Shanghai,200090,China\,\\
\email{chaoswang@foxmail.com}} 
\maketitle

%begin{history}
%received{(Day Month Year)}
%revised{(Day Month Year)}
%accepted{(Day Month Year)}
%comby{(xxxxxxxxxx)}
%end{history}

\begin{abstract}
Exploring proper way to conduct multi-speech feature fusion for cross-corpus speech emotion recognition is crucial as different speech features could provide complementary cues reflecting human emotion status. While most previous approaches only extract a single speech feature for emotion recognition, existing fusion methods such as concatenation, parallel connection, and splicing ignore heterogeneous patterns in the interaction between features and features, resulting in performance of existing systems. In this paper, we propose a novel graph-based fusion method to explicitly model the relationships between every pair of speech features. Specifically, we propose a multi-dimensional edge features learning strategy called Graph-based multi-Feature fusion method for speech emotion recognition. It represents each speech feature as a node and learns multi-dimensional edge features to explicitly describe the relationship between each feature-feature pair in the context of emotion recognition. This way, the learned multi-dimensional edge features encode speech feature-level information from both the vertex and edge dimensions. Our Approach consists of three modules: an Audio Feature Generation(AFG)module, an Audio-Feature Multi-dimensional Edge Feature(AMEF) module and a Speech Emotion Recognition (SER) module. The proposed methodology yielded satisfactory outcomes on the SEWA dataset. Furthermore, the method demonstrated enhanced performance compared to the baseline in the AVEC 2019 Workshop and Challenge. We used data from two cultures as our training and validation sets: two cultures containing German and Hungarian on the SEWA dataset, the CCC scores for German are improved by 17.28\% for arousal and 7.93\% for liking, and for Hungarian, the CCC scores are improved by 11.15\% for arousal and 131.11\% for valence. The outcomes of our methodology demonstrate a 13\% improvement over alternative fusion techniques, including those employing one dimensional edge-based feature fusion approach.

\end{abstract}


\section{Introduction}

Speech is an easily accessible medium for humans to communicate their intentions and messages, which can be conveyed via ubiquitous devices such as microphones and phone. Consequently, many previous studies devoted to recognizing human emotions from their speeches \cite{ref1-chang2021enforcing}. Recent development in speech emotion recognition has greatly transformed consumer shopping, information gathering, and communication \cite{ref2-melumad2023vocalizing}. Practice feedback has been well applied commercially to the fields of human-computer interaction, healthcare, and marketing \cite{ref3-chen2015aiwac,ref4-rautaray2015vision}.

A key challenge in speech emotion recognition is to learn discriminative emotional representations from raw speech signals. Early studies \cite{ref5-xia2020learning} used low-level hand-crafted features  to complete speech emotion recognition tasks, including Low level descriptors(LLDS) and High level statistics functions (HSFs, e.g. mean and maximum values of LLDs) \cite{ref6-liu2021graph}, Mel Frequency Cepstral Coefficients (MFCCs) \cite{ref7-wang2015speech}, Linear Predictive Coding (LPC) \cite{ref8-jagadeeshwar2023asernet}, and Perceptual Linear Prediction (PLP) \cite{ref9-khalil2019speech,ref10-mehra2022beris}. Based on LLDs, researchers applied audio word bag (BoAW) to extract features in an unsupervised manner, and obtained more statistical information by aggregating multiple frame-level descriptive words from audio segments \cite{ref11-zhang2018unsupervised}. However, the above-mentioned hand-crafted features may fail to learn task-specific (e.g., emotion-related) features from the given audio signals. As a result, with the emergence of deep learning model, more and more speech emotion recognition methods have been built on Deep Neural Network (DNNs), which can be optimized based on the target and thus can extract task-specific features from audios to make better predictions \cite{ref12-mcloughlin2015robust,ref13-li2023soil,ref14-ye2021multi,ref15-amiriparian2022deepspectrumlite}. Deep learning methods can extract hierarchical and discriminative feature representations through supervised learning, and the application of models such as CNN and LSTM has also been shown to further improve accuracy \cite{ref6-liu2021graph,ref16-mirsamadi2017automatic,ref17-wang2019speech}. Recently, other artificial intelligence (AI) techniques, such as self-attention mechanisms, have also been proposed for SER tasks \cite{ref18-li2019improved}.

\begin{figure}[!h]
	\centerline{\includegraphics[width=1\linewidth]{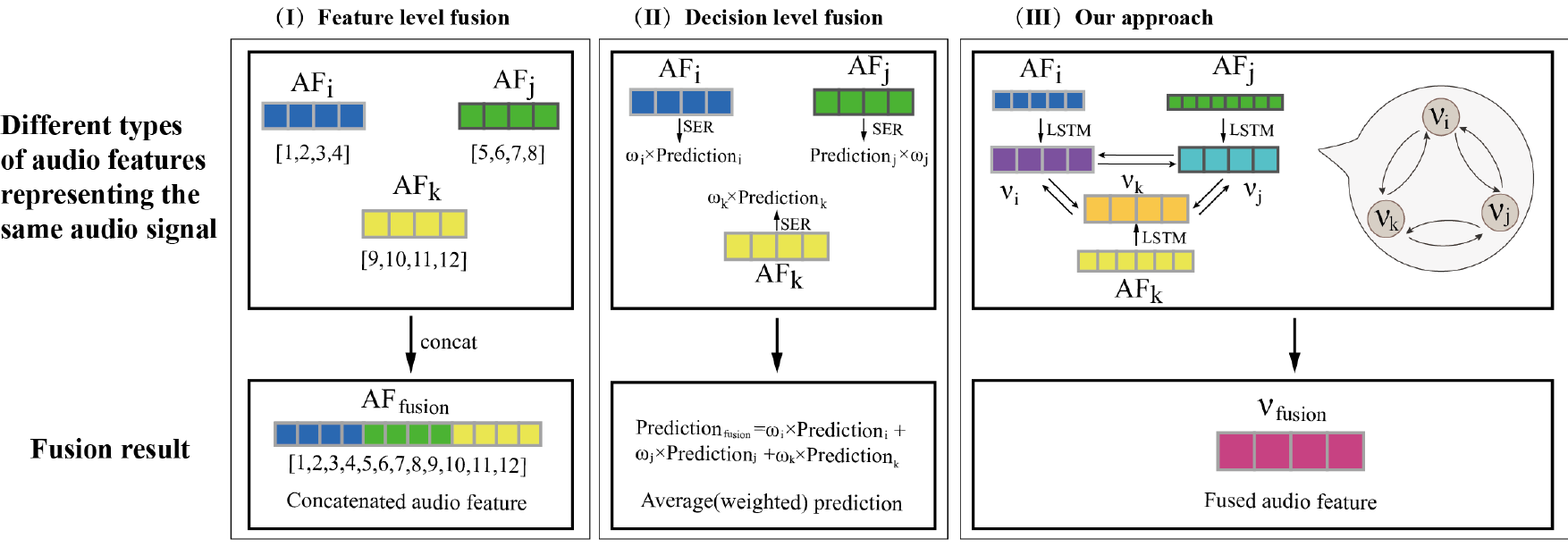}}
	\vspace*{8pt}
	\caption{Comparison between different speech feature fusion methods, with examples shown as follows, Use audio features (AF), recognition results prediction, and feature vectors V. (I) Perform feature-level fusion using concat methods; (II) Perform decision-level feature fusion based on weighted recognition results (I) and (II) without using edge features; and (III) Our method Showing a unique association pattern encoded for each feature vector in the node, these features also determine the specific task topology of the graph, and additionally describe the use of multidimensional features for each edge (the relationship between a pair of AFs).}
\end{figure}

In previous studies, researchers frequently predict emotions based on a single speech feature described above \cite{ref19-nwe2003speech,ref20-jain2020speech,ref21-chuang2004emotion}. In the context of SER tasks, low-level hand-crafted features have been employed in small-sample data sets due to their transparent computational principles and straightforward structure. In contrast, deep learning methods are more suited to feature extraction tasks in massive data sets, as they are capable of automatically learning optimal feature representations and combining low-level features to form novel features \cite{ref22-hashem2023speech,ref23-hossain2019emotion}.
For speech emotion recognition, the use of a single type of feature in some cases leads to low recognition accuracy. First, when dealing with speech data from speakers with different backgrounds, due to the influence of contextual factors such as gender information, language, culture, etc., these problems lead to a single audio feature that cannot cover all the emotional information \cite{ref24-cowie2001emotion}; second, when dealing with speech data with a large amplitude of emotional change, due to the influence of the speaker's emotion, the existence of changes in the speed of speech, as well as the difference in the energy between the various spectra, a single feature can only express speech's emotional information from one side to express the emotional information of speech, so a single type of feature has a performance bottleneck in speech emotion recognition \cite{ref25-liu2021trap,ref26-kim2018emotion}.

To overcome the limitations of speech recognition based on a single speech feature, combining different speech features has proven to be an effective solution \cite{ref27-bandela2023stressed}. Specifically, the most common solution is to directly concatenate multiple different speech features \cite{ref28-paul2023machine}. For example, Zhou et al. combine emotion feature maps extracted from different spectral diagrams, thereby enhancing the accuracy of speech recognition \cite{ref29-guo2023multi}. Alternatively, some studies conduct decision-level fusion, which first make an emotion prediction from each feature and then average all predictions \cite{ref30-akalya2023multimodal,ref31-zhang2023developing}. However, both strategies fail to comprehensively model the inter-relation between different speech features, which may lead to not only the final combined feature contain redundant information but also ignore crucial relationship cues for the emotion recognition. Recently, Graph Neural Networks (GNNs) have attracted more attentions from researchers for speech emotion analysis \cite{ref32-lian2020conversational}. However, the node features used in this study are all deep learning features from LSTM outputs, and as mentioned earlier, a single feature type can provide very limited useful information. Therefore, we propose the hypothesis that using different types of speech features as nodes, more useful information can be learned from neighboring nodes using an effective aggregation method \cite{ref6-liu2021graph}.

In this paper, we propose a novel graph-based fusion strategy that leverages the merits of both deep-learned and hand-crafted audio features for speech emotion recognition. Inspired by GRATIS \cite{ref33-song2022gratis}, we propose a novel method that learns a multi-dimensional edge feature to explicitly describe the relationship between each pair of to address the challenge of feature integration in cross-corpus speech emotion recognition. Compared to methods such as GIN, it has higher coupling with non-graphical heterogeneous data such as deep learning and hand-crafted audio features and can easily be combined with various deep learning backbone and GNN to predict different downstream tasks, facilitating subsequent speech emotion classification. Our approach consists of three modules: an Audio Feature Generation (AFG) module, an Audio-Feature Multi-dimensional Edge Feature (AMEF) module and a Speech Emotion Recognition (SER) module. Initially, it draws upon two conventional features, eGeMAPs and MFCCs, as well as two bag-of-words models, BoAW-e and BoAW-m, and the deep learning feature, Deep Spectrum. AFG module extracts  five feature vectors representing these five speech features within an LSTM framework. In a subsequent phase, the AMEF module was used to construct a graph representation based on multidimensional edges, which is ultimately integrated into a combined feature. Finally, the combined feature is fed back into the model as an input. The SER module completes the emotion recognition of speech vectors through regression methods, resulting in the outcome. The main contribution of this work is summarized as follows:
\begin{itemlist}
    \item To the best of our knowledge, we introduce the first approach to leverage multi-dimensional edge features for graph-based cross-corpus sentiment recognition tasks. By constructing learnable multi-dimensional edge features, we explicitly consider and model the relationships between complex and diverse audio features.
 
    \item We present a method for fusing different types of speech features in a way that captures the information and relationships of different speech features. Through cross-attention, we effectively extract novel features and utilize them to guide the updating of multi-dimensional edge features.
    
    \item We conducted experiments on SEWA dataset,an official cross-lingual speech dataset,significant improvements in the accuracy of speech emotion recognition have been achieved by integrating fusion methods into existing graph-based frameworks.
\end{itemlist}

\section{Related Work}

\subsection{Speech Emotion Recognition Methods}

The primary task of Speech Emotion Recognition (SER) is to extract and identify the emotional information contained in speech.  Over the past several decades, the study of identifying human emotional states from speech through computer technology has been a popular field of research. The SER system's recognition process of speech emotions is divided into two steps.  Firstly, speech features need to be extracted.  These features carry contextual information and are important sources of information for computers to recognize emotions. In early research, the common features used in speech emotion recognition by academia included prosody features, spectral features, and timbre features \cite{ref9-khalil2019speech}. The hand-crafted features include statistical functions such as mean values and extrema, as well as features such as Mel Frequency Cepstral Coefficients (MFCCs), Linear Predictive Coding (LPC), and Perceptual Linear Prediction (PLP) \cite{ref9-khalil2019speech,ref10-mehra2022beris}. On the basis of LLDs, researchers have used the Audio Word Bag (BoAW) to extract features in an unsupervised manner \cite{ref11-zhang2018unsupervised}.

Recently, many studies have begun to use deep learning methods to extract Deep Spectrum representations from audio segments. Deep Spectrum representations are mainly due to the advantages of transfer learning, as they are formed by passing spectral maps through pre-trained image classification CNNs such as AlexNet or VGG \cite{ref34-krizhevsky2012imagenet}. Deep Spectrum features are also typical Deep Spectrum features, and have been proven effective for various general audio recognition tasks \cite{ref35-aghajani2020speech}. For example, the DeepSpectrumLite transfer learning framework is used for pre-training image convolutional neural networks (CNNs) for speech recognition \cite{ref15-amiriparian2022deepspectrumlite}. Moreover, humans may have more than one tone when speaking, and the diversity of tones represents changes in emotion \cite{ref36-begeer2013sex,ref37-al2023speech}. Under such circumstances, the boundaries of emotional recognition become even more complex due to human factors. On the other hand, the perception of speech in a language depends on factors such as the speaker's cultural background, language, gender, and even the speaker's age, which makes speech emotion recognition more complex \cite{ref38-haider2021distributed}. Based on the above description, the second step of speech emotion recognition is to classify the emotional content of the speech file through machine learning and other classification algorithms. In most cases, classifiers are divided into two main categories: traditional classifiers and deep learning classifiers. Traditional classifiers include SVM, HMM, GMM, KNN, decision trees, LDA, and maximum likelihood methods \cite{ref39-SUN2019Speech,ref40-lanjewar2015implementation,ref41-torres2017svm}. In contrast, there has been an increasing number of studies in recent years using deep learning classifiers such as CNN, DNN, RNN, DBN, LSTM, and DBM for speech emotion recognition.  Notably, the emotional content in speech changes over time, so it is appropriate to use techniques that are effective for modeling time, such as stochastic HMMs or neural networks with recursive units, such as LSTM or GRU \cite{ref42-lieskovska2021review}.

Despite notable progress in emotion recognition, cross-corpus speech emotion recognition remains a significant challenge due to cultural and linguistic disparities \cite{ref43-hareli2015cross}. In the CES sub-project of the AVEC 2019 Workshop and Challenge, Fabien and colleagues utilized a 2-layer LSTM-RNN (64/32 units) as the time-dependent regressor for each representation of audio signals and employed SVMs for late-stage fusion of predictions \cite{ref44-ringeval2019avec}. Compared to the performance reported in the previous AVEC CES, arousal and valence scores for German improved by 7.25\% and 8.25\%, respectively, while those for Hungarian improved by 17.3\% and 13.3\%, respectively. However, the study still faces an issue: individual features such as the  MFCCs, eGeMAPs, and deep learning DEEP SPECTRUM cannot achieve optimal results across multiple emotion categories due to varying information content within each feature \cite{ref45-devillers2016geneva}. This is particularly true since these features describe emotional information from different perspectives. Therefore, there is a need to find a method for deep fusion of features to generate new ones, aiming for better results across multiple emotion categories.

\subsection{Graph Representation Learning and Graph Neural Networks}

While deep learning effectively captures the hidden patterns in Euclidean data such as images, text, and videos, an increasing number of applications represent data in graphical form \cite{ref46-sen2008collective}. For instance, in e-commerce, graph-based learning systems can leverage relationships between users and products to make highly accurate recommendations \cite{ref47-wu2020comprehensive}. Graph Neural Networks (GNNs) have garnered significant attention for their ability to handle non-Euclidean data \cite{ref48-ma2019multi}. Thanks to their unique advantages in relationship construction, GNNs can effectively represent multiple features of the same type as nodes in a graph, thereby capturing both similarities and differences between data points \cite{ref49-dai2022graph,ref50-gong2019exploiting,ref51-velivckovic2017graph,ref52-ciano2021inductive}.

The purpose of graph representation learning is to convert nodes in a graph into low-dimensional vector representations through specific methods. This low-dimensional vector representation can preserve the node features, structural features, and semantic information of the original graph, represented by a set of vertices and edges \cite{ref53-isufi2021edgenets,ref54-bengio2013representation,ref55-dwivedi2023benchmarking}. However, most graphs are constructed by manually defining the vertices and the characteristics of each edge through predefined rules, which ignores the links related to the task and affects the performance of graph analysis \cite{ref56-mccallum2000automating}, \cite{ref57-cai2021edge}. As an essential component of the graph, edge features can avoid overlooking crucial feature-relation clues by leveraging the relationships between vertices \cite{ref58-xiong2021multi, ref59-song9993801learning,ref60-shao2021personality}. In recent years, there have been studies using multi-dimensional edge features to describe specific relationships between vertices, such as using heuristic edge features to represent relationships \cite{ref61-wang2020learning}. However, these hand-crafted features still cannot capture specific task relationship information between vertices \cite{ref62-luo2022learning,ref63-hou2022multi}. This requires a graph representation learning framework to automatically generate task-specific topology and multi-dimensional edge features for different input modalities.  Some studies have utilized multiple sub-graphs to learn rich node representations in graph-based networks, proposing two types of edge features and effectively combining them with GAT and GCN models for relationship extraction \cite{ref64-mandya2020contextualised}. Prior to this, multi-dimensional edge features have been widely used in areas \cite{ref65-liu2023smef,ref66-wang2022me,ref67-tao2022revisiting,ref75-xu2023reversible}. Song et al. proposed a general graph representation learning framework (called GRATIS) that can generate strong graph representations with specific task topologies and multi-dimensional edge features for arbitrary inputs. Research results show that GRATIS not only greatly enhances pre-defined graphs but also learns strong graph representations of non-graph data, resulting in significant performance improvements across all tasks \cite{ref33-song2022gratis}. We leverage the GRATIS graph representation learning framework as an auxiliary task, fully utilizing information from various speech feature datasets.

\section{Method}

\begin{figure}[!h]
	\centerline{\includegraphics[width=1\linewidth]{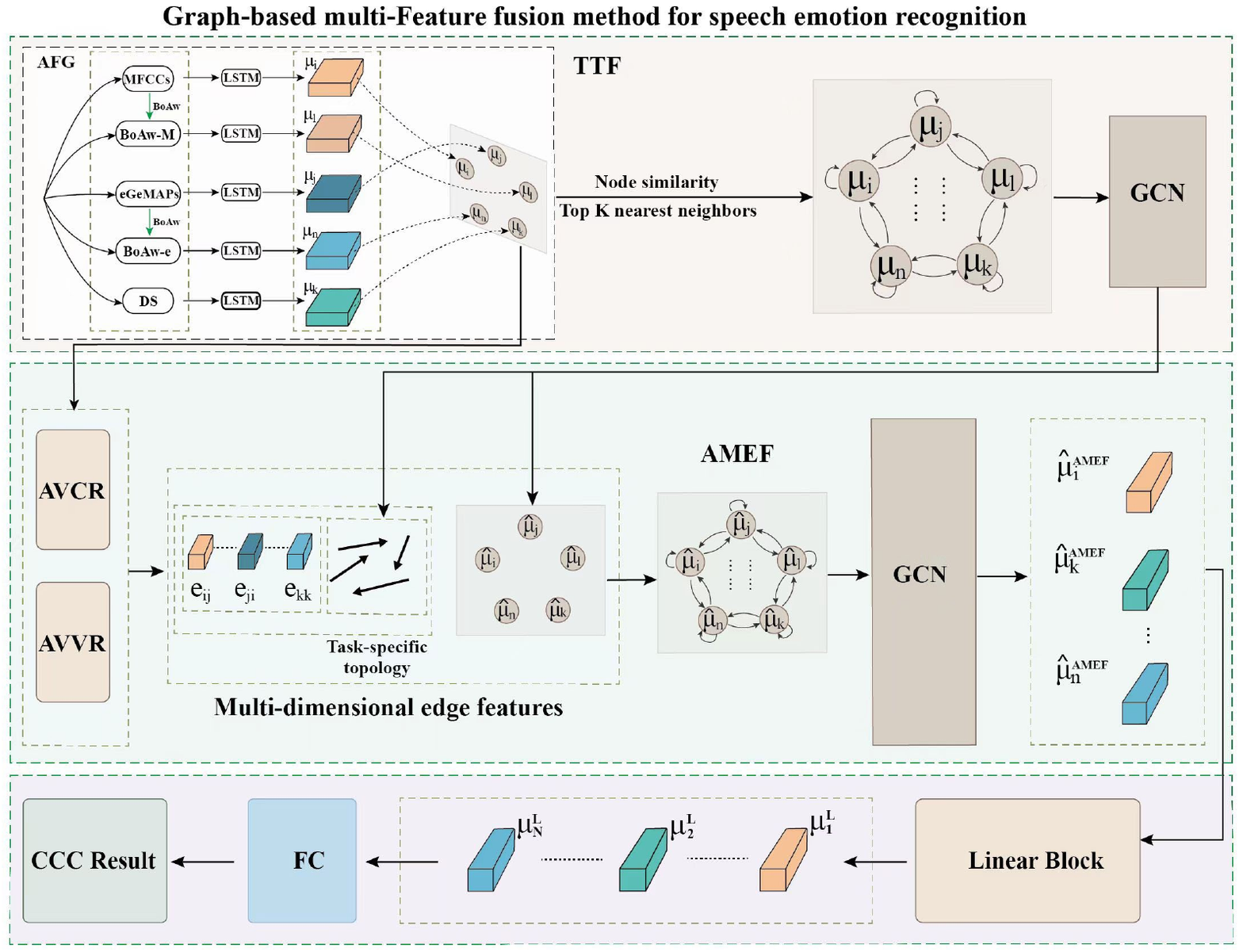}}
	\vspace*{8pt}
	\caption{The framework of the graph-based multi-feature fusion method. For simplicity, only the processing process of an audio file is shown in the graph. The specific details of the modules are given in the corresponding sections.}
\end{figure}

In this section, we propose a novel graph-based speech feature fusion method, where multi-dimensional edge features are learned to specifically describe the relationship between every pair of speech features. 

\textbf{Overview:} The overall architecture is illustrated in Figure 2. First, our Audio Feature Generation (AFG) module extracts features from the given audio signal, including five hand-crafted features, namely eGeMAPs and MFCCs, two bag-of-words model features, BoAW-e, and BoAW-m, as well as deep learning features, Deep Spectrum. These five features encompass different speech emotional information. The feature extraction is accomplished through two layers of LSTM. Secondly, Due to the varying dimensions of diverse types of speech features, a 2-layer LSTM-RNN (64/32 units) can be employed to achieve a uniform feature vector dimensionality. Thirdly, the GRATIS framework generate graph representations with task-specific topology and multidimensional edge features, since speech features are non-graph data, the input speech features will be represented by a set of vectors and the GRATIS framework connects them into a matrix, by training to obtain the optimal topology. This vector, through edge feature information, contains the association between features, and through point feature information, represents the differences between features. Finally, the vector is fed back into the model as an input, and the end-to-end output is obtained through a fully connected layer.

\subsection{Hand-crafted and Deep learning speech features extraction}

Identifying emotions from audio-visual signals usually relies on feature sets, which are extracted based on different aspects that help the model acquire knowledge. In the first module of Figure 2 (AFG), the study achieve the extraction of handcrafted speech features.

\begin{itemlist}
	\item {eGeMAPs features} using the extended Geneva Minimum Acoustic Parameter Set (eGeMAPs). This set encompasses 88 metrics covering the acoustic dimensions, include spectral, cepstral, prosodic, and speech quality information.
	\item {MFCCs features} have 13 coefficients, which have better robustness than LPCCs based on vocal tract models, and have been proven to have good recognition performance even when the signal-to-noise ratio decreases \cite{ref68-murty2006combining}. And the extraction of MFCCs was accomplished using the openSMILE2 toolkit \cite{ref45-devillers2016geneva}, \cite{ref69-eyben2013recent}.
	\item {Bag-of-words-eGeMAPs (BoAW-e) features} are based on the bag-of-words model features of handcrafted features, eGeMAPs LLDs are standardized in an online manner (zero mean, unit variance) before vector quantization. The entire cross-modal BoW (XBoW) processing is executed using the open-source toolkit openXBOW4 chain \cite{ref70-ringeval2018avec}.
    \item {Bag-of-words-MFCCs (BoAW-m) features} are based on the bag-of-words model features of handcrafted features, MFCCs LLDs are standardized in an online manner before vector quantization. The entire cross-modal BoW (XBoW) processing is executed using the open-source toolkit openXBOW4 chain \cite{ref70-ringeval2018avec}.
	\item {Deep Spectrum features}, the inspiration for Deep Spectrum features comes from image processing. By inputting the spectral image of a speech instance into a pre-trained image recognition CNN and extracting a set of resulting activation values as feature vectors. Deep Spectrum features have been widely applied in past speech emotion recognition research for their efficient extraction ability and good recognition preparation rate \cite{ref70-ringeval2018avec}, \cite{ref71-amiriparian2017snore}.We employed the VGG-16 to extract Deep Spectrum features from the pre-trained CNN \cite{ref72-simonyan2014very}. A 4096-dimensional feature vector is then extracted from the activations of the second fully connected layer in VGG-16.
\end{itemlist}

In summary, a total of five different features can be obtained, the LLDs features contain acoustic features with a small number of dimensions, which can effectively reduce the computational complexity of the subsequent pattern recognition system; whereas the Deep Spectrum features contain more dimensions, which have high computational complexity when inputted to the recognition system but contain spatio-temporal features, such as temporal information, and each has its own advantages and disadvantages. Therefore, we expect to further improve their performance in speech emotion recognition scenarios through feature fusion.

\subsection{Audio-feature temporal modelling}

Before feature fusion, we need to preprocess different types of features, The method comprises two stages. The first stage involves unifying the vector dimensions of each type of speech feature, represented as a graph comprising speech features with the same vector dimensions. This reduces the computational complexity of the subsequent speech emotion recognition model. Second, to consider the temporal sequence when processing the feature vectors, due to the time-series nature of the speech data, the LSTM is more time-sensitive than the ordinary RNN, and it can learn the patterns and features in the time-series data. Therefore, we choose a 2-layer LSTM-RNN to standardize the dimension of speech feature vectors. The 2-layer LSTM layers with different units can better capture the key information between sequential data.

\subsection{Audio feature fusion method}

\subsubsection{Vertex Feature Extraction}

Five different speech features (eGeMAPs, MFCCs, BoAW-M, BoAW-e and DS) are obtained from the input audio feature data,these five speech features are used as the vertex features with the $\mu$.  Firstly, The graph representation consists of five vertex features, each representing a speech feature: the hand-crafted features eGeMAPs and MFCCs, the two bag of words features BoAW-e and BoAW-m, and Deep Spectrum features. Each speech feature is processed by two LSTM layers in order to obtain a feature vector $\mu$ with a unified dimension of $[N, K]$, the representative speech data has $N$ frames and each feature contains $K$ dimensions. Then, a set of vertex features for each frame can be represented as:
\begin{equation}
\mu=\left(\mu_i,\mu_j,\mu_k,\mu_n,\mu_l\right)
\end{equation}
where the vertex ${\mu}_i$ represents the use of eGeMAPs features processed into feature vectors by two-layer LSTM; The vertex ${\mu}_j$ represents the use of MFCCs features processed into feature vectors by two-layer LSTM; The vertex ${\mu}_k$ represents the use of BoAW-M features processed into feature vectors by two-layer LSTM; The vertex ${\mu}_n$ represents the use of BoAW-e features processed into feature vectors by two-layer LSTM; The vertex ${\mu}_l$ represents the use of Deep Spectrum features processed into feature vectors by two-layer LSTM.

The vertex features are initially fed to a backbone, which may be any suitable machine learning model. In the case of graph data, this may be a GNN, whereas in the case of non-graph data, this may be a Transformer. In this study, we have chosen to use a BiGRU model as the backbone. The composition of the backbone is due to the fact that the BiGRU model introduces a bi-directional structure on top of the GRU model, which better captures the bi-directional dependencies of sequential data, such as speech features. In our implementation, we use GRU as the backbone, to directly extract the global contextual representation$\mathbf{\/Y}\in{\mathbb{R}}^{A\times K\times N}$   from the input data, where A denotes the number of vertices. The global contextual representation $Y$ is defined as:
\begin{equation}
	Y=BiGRU{\left(\mu\right)}
\end{equation}

\subsubsection{Graph Definition}

We need to use the vertex features $\mu$, the basic adjacency matrix $\mathcal{A}$ and edge features $\beta$ to build its graph representation $\mathcal{G}^A(\mu,\beta)$. The number of vertex feature extractors is $i$, each of which consists of a fully connected layer (FC). The vertices set $\mu$ contains vectors describing five audio features (each has $K$ dimensions). We treat each vector directly as a vertex, and the basic adjacency matrix concatenating all vectors as the global contextual representation $Y$. We define edge presence in the basic graph $\mathcal{G}^A$ according to $\mathcal{R}ule$, These can be formulated as:
\begin{equation}
	\begin{matrix}
	    \mathcal{G}=\mathcal{G}^A(\mu,\beta)
\\\beta\subseteq\left\{\mathbf{e}_{i,j}=1\mid\mathbf{\mu}_i,\mathbf{\mu}_j\in\mu\mathrm{\ and\ } \mathcal{A}_{i,j}=1\right\}\\\end{matrix}
\end{equation}
which is subject to:
\begin{equation}
	\mathcal{A}_{i,j}=\left\{\begin{matrix}1&\left\{\mathbf{\mu}_i,\mathbf{\mu}_j\right\}\in\mathcal{R}ule\\0&\mathrm{\ Otherwise\ }\\\end{matrix}\right.
\end{equation}
where $\mathbf{e}_{i,j}$ represents the vertex ${\mu}_i$ effect on its adjacent vertex ${\mu}_j$ is controlled  by a single value. $\mathcal{R}ule$ represents the basic edge representation between a pair of connected vertex $i$ and vertex $j$ is defined as 1.

\subsubsection{The task-specific topology and vertex features(TTF)}
This section takes the global contextual representation $Y$, the basic adjacency matrix $\mathcal{A}$ and vertex features ${\mu}$ as the input, and generates the task-specific adjacency matrix topology and vertex features.

Firstly,  we consider the feature vectors as five node features and define the connectivity (vertex-to-vertex relationship) and the connectivity between a pair of nodes $\boldsymbol{\mu}_i$ and $\boldsymbol{\mu}_j$  by the similarity of the features($Sim(i,j)= \boldsymbol{\mu}_i^T \boldsymbol{\mu}_j$). At the same time, we select the K nearest neighbors of each node as its neighbors, thus completing the definition of the graph topology. Then, a GCN layer is employed to jointly update all vertexs from the produced graph, where the $i_{th}$  node's task-specific vertex feature $\boldsymbol{\hat{\mathcal{\mu}_i}}$ is generated by $\boldsymbol{\mu}_i$ and its connected nodes as:
\begin{equation}
\boldsymbol{\hat{\mathcal{\mu}_i}}=Relu\left[\boldsymbol{\mu}_i + p\left(\boldsymbol{\mu}_i,\sum_{j=1}^{A} q\left(\boldsymbol{\mu}_j, a_{i,j}\right)\right)\right]
\end{equation}
where ${Relu}$ is the nonlinear activation function, ${p}$ and ${q}$ denote functions of the GCN layer, ${a_{i,j}}$ represents the relationship between two vertices. 

After obtaining the optimal task-specific vertex feature representation, we then use this to calculate the vertex-to-vertex similarity to obtain the task-specific adjacency
matrix topology $\mathcal{\hat{A}}_{i,j}$.
\begin{equation}
\mathcal{\hat{A}}_{i,j}=\boldsymbol{\hat{\mathcal{\mu}_i}} \boldsymbol{\hat{\mu}_j^T}
\end{equation}

\subsubsection{Audio-Feature Multi-dimensional Edge Feature Learning (AMEF)}

To further learn various cues that describe the relationship between speech features and specific tasks, each learned edge is a 1$\times$N dimension, which assign a multi-dimensional to each learned edge feature ${\hat{\mathbf{e}}}_{i,j}\in\mathbb{R}^{1\times N}$. This module is trained under the supervision of target graph analysis tasks. Due to the complex information dimension of speech features, including mathematical information and temporal information, we assume that the relationship between the connected multi-dimensional vertices $\hat{\mathbf{\mu}}_i$ and $\hat{\mathbf{\mu}}_j$ cannot be best described by one-dimensional edge features. In addition, this relationship clue is not only contained in their vertex features, but also reflected in the global context of the graph. The fusion method takes various speech features $\mu$ and global context representation $Y$ as inputs and outputs a multi-dimensional edge feature vector.

The Audio-feature Vertex-Context Relationship Modelling (AVCR) block takes vertex features $\hat{\mathbf{\mu}}_i$ and $\hat{\mathbf{\mu}}_j$ and the global contextual representation $Y$ as input. It conducts cross attention between $\hat{\mathbf{\mu}}_i$ and $Y$ as well as $\hat{\mathbf{\mu}}_j$ and $Y$. Before that, the global contextual representation $Y$ goes through a reshape operation into $\hat{Y}$ to keep the shape consistent with that of $\hat{\mathbf{\mu}}_i$ and $\hat{\mathbf{\mu}}_j$. The vertex features $\hat{\mathbf{\mu}}_i$ and $\hat{\mathbf{\mu}}_j$ are independently used as queries to locate vertex-context relationship features $\mathcal{F}_{i,x}$ and $\mathcal{F}_{j,x}$ in $\hat{Y}$. Mathematically speaking, this process can be represented as:
\begin{equation}
	\begin{matrix}
		\mathcal{F}_{i,x}=AVCR(\hat{\mathbf{\mu}}_i,\hat{Y}) \\
		\mathcal{F}_{j,x}=AVCR(\hat{\mathbf{\mu}}_j,\hat{Y})
	\end{matrix}
\end{equation}
with the cross-attention operation in AFR defined as:
\begin{equation}
	AVCR{\left(A,B\right)}=softmax{\left(\frac{AS_q\left(BS_k\right)^T}{\sqrt{f_k}}\right)}BS_\mu
\end{equation}
where ${S_q}$,${S_k}$,${S_\mu}$ are learnable weight  matrices for the encoding of the query, the key and the values, and ${f_k}$ is a scaling factor.

The Audio-feature Vertex-Vertex Relationship Modelling (AVVR) block further extracts the features. It also conducts the cross-attention (has the same form as Eq. 10 but independent weights) produces features $\mathcal{F}_{i,j,x}^{}$ and $\mathcal{F}_{j,i,x}^{}$, where $\mathcal{F}_{i,j,x}^{}$ is generated by using $\mathcal{F}_{j,x}^{}$ as the query and $\mathcal{F}_{i,x}^{}$ as the key and value. Finally, we feed $\mathcal{F}_{i,j,x}^{}$ and $\mathcal{F}_{j,i,x}^{}$ to a fully-connected (FC) layer to obtain multi-dimensional edge feature vectors $\boldsymbol{e}_{i,j}$ and $\boldsymbol{e}_{j,i}$, respectively. This process can be represented as:
\begin{equation}
	\boldsymbol{e}_{i,j}, \boldsymbol{e}_{j,i}=FC\left(AVVR\left(\mathcal{F}_{j,x}^{}, \mathcal{F}_{i,x}^{}\right), AVVR\left(\mathcal{F}_{i,x}^{}, \mathcal{F}_{j,x}^{}\right)\right)
\end{equation}
In order to better integrate the multidimensional edge feature learning method for multiple speech features, The graph was constructed using the GCN model after the multidimensional edge features were obtained. These features were then combined with the previously obtained task-specific topology and vertex features. The task-specific vertex features were used to create the task-specific topology, which was then used as input.The task-specific topology is unified with the multidimensional edge feature through multiplication. Prior to this, the task-specific topology is reshaped once, with the reshaped dimension matching that of the multidimensional edge features.
\begin{equation}
	\boldsymbol{\hat{e}}=\boldsymbol{{e}} \mathcal{\hat{A}}
\end{equation}
\begin{equation}
    \mathbf{\hat{\mu}^{AMEF}},\mathbf{\hat{e}^{AMEF}}=\mathbf{GCN}\left(\mathbf{\hat{\mu}},\mathcal{\boldsymbol{\hat{e}}}\right)
\end{equation}
where $\hat{\mu}$ represents the set of all task-specific vertex features obtained in the TTF section, $\mathcal{\hat{A}}$ represents the task-specific topology between all vertices.

\subsection{Feature fusion and emotion recognition}

Once the audio-feature relation graph $\boldsymbol{G}^L=\left(\boldsymbol{\mu}^L, \boldsymbol{e}^L\right)$ that consists of $A$ node features and $A\times A$ multi-dimensional directed edge features is learned, we feed this into the GCN model to jointly recognise all audio data. The next step is the feature fusion, where graphs containing multiple classes of features are fed to through the employed GCN network, and then a final fusion vector is obtained. Specifically, in the reasoning phase, the graph consists of multiple types of speech features, using the number of nodes to determine the structure and size of the output layer, this leads to feature fusion. In initialising the weights, the number of nodes $Num$ is used to initialise the weight matrix. In the step of batch normalisation layer operation, the layers normalised against node features, their dimension is set to $Num$ and in the layers normalised against edge features, their dimension is set to $Num \times Num$ as the feature matrix of the edges is two dimensional.  In the update step of the nodes, $Num$ is used to normalise the aggregated edge features, i.e. divide by $Num$. This is because when calculating the update for each node, it is necessary to take into account how many edges point to that node. Compared to traditional feature fusion methods, our method first fuse node features by applying convolution operation on the graph structure, through which not only the features of the node itself are fused, but also the features of the neighbouring nodes are taken into account, and then the edge information is preserved by the method of multidimensional edge features. Secondly, the output layer of our method usually uses a fully connected layer whose output dimension is equal to $Num$, and then, the features of each node are mapped to different class probabilities. Finally, in contrast to approaches that perform fusion at the feature level and fusion at the decision level, our approach allows dynamic feature fusion at different levels of the graph structure, which provides more flexibility to the model.

After completing the above updates, we obtained the final output of the AMEF module $\boldsymbol{\mu}^L$, which is a $N \times Num$ feature vector, where N is the dimension of the audio data and $Num$ represents the number of classifications set by the model. At the end of the model, we add a linear layer with the purpose of transforming $Num$ in the output of the AMEF module into a feature vector with a value of 1.This step is the fusion of speech features, we fuse five features into a new feature vector and use the feature vector to complete the task of speech emotion recognition.

\subsection{Training scheme}

The training set and validation set are divided according to cultural types, where the training set is divided into monolingual data sets and multilingual data sets. We use two types of data sets to complete the experiment, and input the fused feature vectors into the 2-layer LSTM-RNN. For time series prediction problems, it is equivalent to setting sequence length to 1. The prediction results are processed by squeeze dimensionality reduction, performing standardization of the prediction results $r_i$ and label values $r_j$. Calculate the mean and variance of the output prediction values and label values. As shown in equations (12) and (13):
\begin{equation}
	\bar{r}=\frac{\sum_{n=1}^N r}{N}
\end{equation}
\begin{equation}
	r\_val=\frac{\sum_{n=1}^N (r-\bar{r})^2}{N}
\end{equation}
where prediction results is $r_i$ and label values is $r_j$, The N represents the number of data points.
The correlation coefficient $\sigma_{r_ir_j}$ is calculated as follows:
\begin{equation}
	\sigma_{r_ir_j}=\left(r_i-{\bar{r}}_i\right)\cdot\left(r_j-{\bar{r}}_j\right)
\end{equation}
The loss function for training and validation is based on the consistency correlation coefficient (CCC) as shown below:
\begin{equation}
	ccc\_loss=C-\frac{2\cdot\sigma_{r_ir_j}}{\left({\bar{r}}_i-{\bar{r}}_j\right)^2+r_{i_-var}+r_{j_-var}}
\end{equation}
where $C$ represents a constant, and in this study, and $C$ is taken as 2 when calculating CCC loss.
The objective of this loss function is to update the model in order to identify the model with the highest CCC score.

\section{Experiment}

This section introduces the experimental details and results. Section 4.1 first presents the dataset. Then, Sections 4.2 and 4.3 introduce Evaluation Metrics and Implementation Details, respectively. Subsection 4.4 explains the advantages of our method compared to the baseline.  Finally, subsection 4.5 shows ablation experiments and analyze the accuracy of speech emotion recognition with different fusion features and our fusion features.

\subsection{Dataset}

We conducted extensive experiments on a cross-cultural speech dataset: the SEWA dataset, which was officially authorized for access and use \cite{ref74-kossaifi2019sewa}. The SEWA database comprises 199 experimental sessions, involving 398 participants from six distinct cultures (UK, Germany, Hungary, Greece, Serbia, and China). We selected the SEWA data used in the AVEC 2019 Workshop and Challenge, encompassing the German and Hungarian cultures, the emotional dimensions arousal and valence, and liking, were independently assessed by several native speakers  \cite{ref44-ringeval2019avec}. 
\begin{table}[!h]
	\tbl{Number of subjects and duration of the video chats contained in the SEWA database \protect\cite{ref74-kossaifi2019sewa}.}
	{\begin{tabular}{@{}cccc@{}} \toprule
			Culture	  & 	Partition	  & 	\# Subjects	  & 	Duration [h:min:s]	  \\
			\colrule
			German	  & 	Training	  & 	34	  & 	1:33:12	  \\
			German	  & 	Devel	  & 	14	  & 	0:37:46	  \\
			Hungarian	  & 	Training	  & 	34	  & 	1:08:24	  \\
			Hungarian	  & 	Devel	  & 	14	  & 	0:28:42	  \\
			All	  & 		  & 	96	  & 	3:48:04	  \\
			 \botrule
	\end{tabular}}
\end{table}

\subsection{Evaluation Metrics}

To ensure comparable results, we used a training and validation set consistent with Ringeval et al \cite{ref44-ringeval2019avec}. These datasets were taken from the SEWA dataset, and the training set contained both German and Hungarian cultures, and for training and validation we used three combinations of data: data from the German culture only, data from the Hungarian culture only, and data from both the German and Hungarian cultures for training. cultures, as well as adding new cultures to the test set: China and the UK, in order to evaluate the potential of our approach in cross-cultural speech emotion recognition tasks. In all experiments, we evaluated our model and baseline using a widely adopted evaluation protocol. The consistent correlation coefficient (CCC) was used as the evaluation metric. as the evaluation metric.

\subsection{Implementation Details}

The operating system we used in the experimental environment configuration is Ubuntu 18.04, the development language is python3.9.0, the framework is Pytorch1.11.0+CUDA 11.1, the CPU is AMD EPYC 7742 64-Core Processor, and the GPU is NVIDIA A100-SXM4-40GB. The experimental part is divided into three parts: feature extraction, feature fusion, and emotion recognition. In the detail section of the feature extraction process, the settings proposed by Ringeval et al. were employed in conjunction with the experimental apparatus \cite{ref44-ringeval2019avec}.We adopted a 2-layer LSTM-RNN (64/32 units) as the time-dependent regressor. We utilized the RMSprop optimizer, implementing a 10\% dropout rate over 50 epochs. The model parameter settings were aligned with the literature \cite{ref70-ringeval2018avec}. And the detailed part of feature fusion and emotion recognition, the network configuration and learning settings of the proposed method are as follows:

\begin{itemize}
	\item GRU is used as the backbone network to extract features from the input graph representation network. GCN is updated during the end-to-end training of the entire system.
	\item The end-to-end training of the entire network, and batch normalization is applied to each fully connected (FC) layer. RMSprop is used as the optimizer, the initial parameter learning rate was set at 0.005 and after obtaining the experimental results, it was subsequently increased by 0.001 for each experiment up to 0.01. This process was repeated a number of times to determine the best results.
	\item The final classifier consists of a fully connected layer.
	\item The method we propose is trained in an end-to-end manner, using the fused feature vectors as inputs.
\end{itemize}

\subsection{Comparison with existing approaches}
We examined the performance of the following model. The baseline algorithm primarily draws inspiration from the studies by Ringeval et al, MFCCs, eGeMAPS, BoAW-M, BoAW-e and DS \cite{ref44-ringeval2019avec}. The reason for using the above results as baseline is that the experimental results of this study are among the better ones in the speech emotion recognition literature using the SEWA dataset in the last few years \cite{Ringeval2017AVEC2R,10.1145/3133944.3133952}.

\begin{table}[!h]
	\tbl{Baseline results evaluated with CCC for the AVEC 2019 CES(MFCCs, eGeMAPs, BoAW-M, BoAW-e and DS)\protect\cite{ref44-ringeval2019avec}; Our approach; best result on the test partition is highlighted in bold.}
	{\begin{tabular}{@{}ccccccc@{}} \toprule
			 Culture	 & 	MFCCs\cite{7160715}	 & 	eGeMAPS\cite{10.1145/2502081.2502224}	 & 	BoAW-M \cite{2016arXiv160506778S}	 & 	BoAW-e	\cite{2016arXiv160506778S} & 	DS \cite{20ff0d256c3441fb90cfb58882bfba1b}		 & 	Ours	 \\
			\colrule
			 & 	\multicolumn{6}{c}{\textbf{\emph{Arousal}}} 		 \\
			 DE+HU	 & 	0.326 	 & 	0.371 	 & 	0.298 	 & 	0.398 	 & 	0.312 	 & 	 \textbf{0.408} 	 \\
			 DE	 & 	0.389 	 & 	0.396 	 & 	0.323 	 & 	0.434 	 & 	0.380 	 & 	\textbf{0.509} 	 \\
			 HU	 & 	0.236 	 & 	0.305 	 & 	0.237 	 & 	0.291 	 & 	0.156 	 & 	\textbf{0.339} 	 \\
			 & 	\multicolumn{6}{c}{\textbf{\emph{Valence}}}		 \\
			 DE+HU	 & 	0.187 	 & 	0.286 	 & 	0.134 	 & 	\textbf{0.352} 	 & 	0.233 	 & 	0.327 	 \\
			 DE	 & 	0.344 	 & 	0.405 	 & 	0.190 	 & 	0.455 	 & 	0.317 	 & 	\textbf{0.460} 	 \\
			 HU	 & 	0.017 	 & 	0.073 	 & 	0.042 	 & 	0.135 	 & 	0.084 	 & 	\textbf{0.312} 	 \\
			 & 	\multicolumn{6}{c}{\textbf{\emph{Liking}}} 		 \\
			 DE+HU	 & 	0.144 	 & 	0.159 	 & 	0.074 	 & 	0.138 	 & 	0.142 	 & 	\textbf{0.176} 	 \\
			 DE	 & 	0.159 	 & 	0.136 	 & 	0.140 	 & 	0.003 	 & 	0.164 	 & 	\textbf{0.177} 	 \\
			 HU	 & 	0.115 	 & 	0.192 	 & 	-0.027 	 & 	\textbf{0.253} 	 & 	0.121 	 & 	0.240 	 \\
			\botrule
	\end{tabular}}
\end{table}

Table 2 shows the overall performance compared to the baseline. From the results, we observe that the proposed model consistently outperforms all the baselines, indicating the effectiveness of the fusion method in speech feature fusion. More precisely, the method proposed in this paper improves upon the baseline in the AVEC 2019 Workshop and Challenge: on the SEWA dataset, the CCC scores for German are improved by 17.28\% for arousal and 7.93\% for liking, and for Hungarian, the CCC scores are improved by 11.15\% for arousal and 131.11\% for valence. It is noteworthy that the efficacy of our method is less pronounced in the 'DE+HU' dataset, with a mere 2.5\% improvement observed in Arousal and a more pronounced decline in Valence compared to the previous baseline. This may be attributed to the existence of significant linguistic differences between languages, which are not adequately captured by the multidimensional edge features. Consequently, the resulting feature representation may be less accurate. This indicates that by integrating speech features through GRU and multi-dimensional edge feature learning, potential feature information can be extracted, thereby enhancing prediction accuracy.

\subsection{Ablation analysis}
%%% 你下面做了几组不同的分析，建议每段用 \textbf{XXX} 订个小标题，这样更加清晰

Additionally, a series of ablation studies were conducted to investigate various aspects of the proposed approach. Five distinct methods for fusing audio features were considered, as illustrated in the ablation model below:

\begin{arabiclist}
	\item   A fusion model based on five audio features, in which the feature vectors are averaged pairwise (Feature fusion-AVE).
	\item Feature Fusion (FC) using fully connected networks (Feature fusion-FC)
	\item  Only use one-dimensional edge to complete the construction of the graph, and perform feature fusion in vector dimensions (GNN+OD-edge)
	\item  Use vector stitching to complete feature fusion, and the output dimension of the vector changes from $N \times K$ to $N \times (K*5)$(GNN+ST).
	\item Based on the five types of audio features, the output prediction values are averaged to complete the calculation of the result. (Baseline-ave).
\end{arabiclist}

\begin{figure}[!h]
	\centerline{\includegraphics[width=0.8\linewidth]{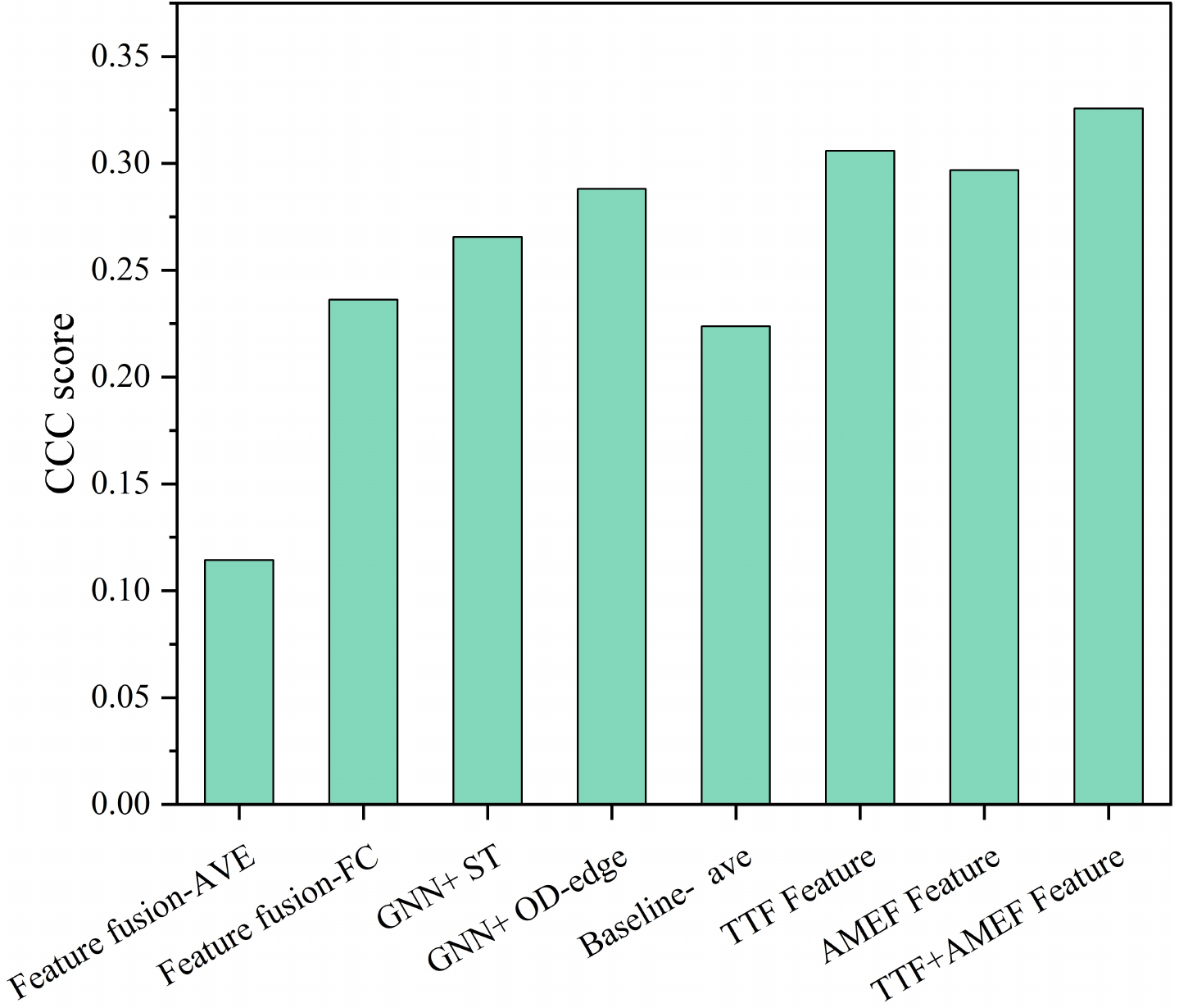}}
	\vspace*{8pt}
	\caption{The values on the vertical axis are the average CCC scores of the three emotional dimensions: arousal, valence, and linking. The horizontal axis represents the ablation model and our method.}
\end{figure}

To more accurately assess the extent to which the different modules influence the final result, three distinct data sources were employed for the purpose of speech emotion recognition. The first approach utilises solely the task-specific vertex features derived from the TTF part for linear layer processing, thereby completing the prediction. The second approach employs the original vertex features that have not undergone processing by the TTF module, in conjunction with the initial topology, to complete the AMEF module. Additionally, the new feature vectors derived from the aforementioned training are subjected to linear layer processing, thereby completing the prediction. The training data are subjected to linear layer processing and complete the prediction. The third method is based on the aforementioned complete method and uses the task-specific topology and vertex features as inputs. The new feature vectors obtained after training through the AMEF module are processed in linear layers and complete the prediction.

\textbf{CCC Score results:} As shown in the Table 3, the results indicate that the fusion method consistently outperforms the models proposed in ablation studies in terms of CCC Scores. The fusion method further improves the fusion of all speech features, most notably by accomplishing dimensionality unification, and the new features do not increase the computational complexity of the emotion recognition model as deep spectral features do. To compare data more intuitively, this study averaged the three emotion classification results (Arousal, Valence, and Liking) of the "DE+HU" cross-cultural dataset. 
\begin{table}[!h]
	\tbl{Ablation analysis results evaluated with CCC Score.}
	{\begin{tabular}{@{}ccccccccc@{}} \toprule
			\multirow{2}{*}{Culture}	 	 & 	Feature	 & 	Feature	 & 	GNN+	 & 	GNN+	 & 	Baseline- & TTF	 & 	AMEF & TTF+AMEF 	 \\
			& 		fusion-AVE	 & 	fusion-FC	 & 	ST	 & 	OD-edge	 & 	ave	 & Feature  & 	Feature & Feature	 \\
			\colrule
			& 	\multicolumn{8}{c}{\textbf{\emph{Arousal}}}		 \\
			DE+HU	  	 & 	0.169 	 & 	0.361 	 & 	0.347 	 & 	0.361 	 & 	0.341 	 & 	0.374 & 	0.372 & 	\textbf{0.408}	 \\
			DE	  	 & 	0.270 	 & 	0.398 	 & 	0.476 	 & 	0.466 	 & 	0.384 	 & 	0.483 & 	0.491 & 	\textbf{0.509}	 \\
			HU	  	 & 	0.045 	 & 	0.308 	 & 	0.291 	 & 	0.293 	 & 	0.257 	 & 	0.324 & 	0.319 & 	\textbf{0.339} 	 \\
			& 	\multicolumn{8}{c}{\textbf{\emph{Valence}}} 		 \\
			DE+HU	  	 & 	0.144 	 & 	0.273 	 & 	0.292 	 & 	0.328 	 & 	0.238 	 & 	0.327  & 	0.323 & 	\textbf{0.330}	 \\
			DE	  	 & 	0.204 	 & 	0.416 	 & 	0.409 	 & 	0.388 	 & 	0.342 	 & 	0.436 & 	0.436 & 	\textbf{0.460}	 \\
			HU	  	 & 	0.053 	 & 	0.129 	 & 	0.300 	 & 	0.257 	 & 	0.070 	 & 	0.306 & 	0.278 & 	\textbf{0.312}	 \\
			& 	\multicolumn{8}{c}{\textbf{\emph{Liking}}} 		 \\
			DE+HU	  	 & 	0.032 	 & 	0.097 	 & 	0.074 	 & 	0.167 	 & 	0.131 	 & 	0.124 & 	0.127 & 	\textbf{0.176} 	 \\
			DE	 	 & 	0.004 	 & 	0.086	 & 	0.077 	 & 	0.136 	 & 	0.120 	 & 	0.139  & 	0.131 & 	\textbf{0.177}	 \\
			HU	  	 & 	0.108 	 & 	0.058 	 & 	0.124 	 & 	0.196 	 & 	0.131 	 & 	\textbf{0.240}  & 	0.195 & 	0.220	 \\
			\botrule
	\end{tabular}}
\end{table}

As shown in Figure 3, based on the average value, it can be found that our fusion method is 13\% better than the results GNN+OD-edge methods. This also reflects that these simple series and parallel methods result in a decrease in model accuracy due to the neglect of edge features. The GNN+ST method and GNN+OD-edge method has considered the importance of edge features, so their results are better than the previously mentioned Feature fusion-FC method that only considers point features. However, our multi-dimensional edge feature fusion results are still better than these single-dimensional edge feature fusion methods, this could be because multi-dimensional edge features can express key information between different speech features, highlighting the role of vertex and edge relationship information in speech emotion recognition.

\textbf{t-SNE results:} As shown in Figure 4, to demonstrate the efficacy of our method, we utilize t-SNE to illustrate the impact of the aforementioned approach in the context of an ablation experiment. To this end, we employ a dataset comprising 40,032 frames of speech belonging to the "DE+HU" group. Each data point is represented by a 10-dimensional output vector, which is subsequently transformed into two dimensions through the use of a dimensionality reduction tool. This transformation allows us to represent the horizontal and vertical coordinates of the points in the graph, respectively. The color of the point is represented by the label value of the point's sentiment category (arousal, validity, and liking). The larger the value, the redder the color, and the smaller the value, the bluer the color. As can be seen from the figure, our method has better clustering results, as the darker points (indicating strong emotions) have higher overlapping coordinates in the plot. In contrast, for methods with suboptimal experimental results, such as the “Feature fusion-AVE” group, the distribution of points with strong emotions is more dispersed in the t-SNE plots of arousal and validity.

\begin{figure}[!h]
	\centerline{\includegraphics[width=1\linewidth]{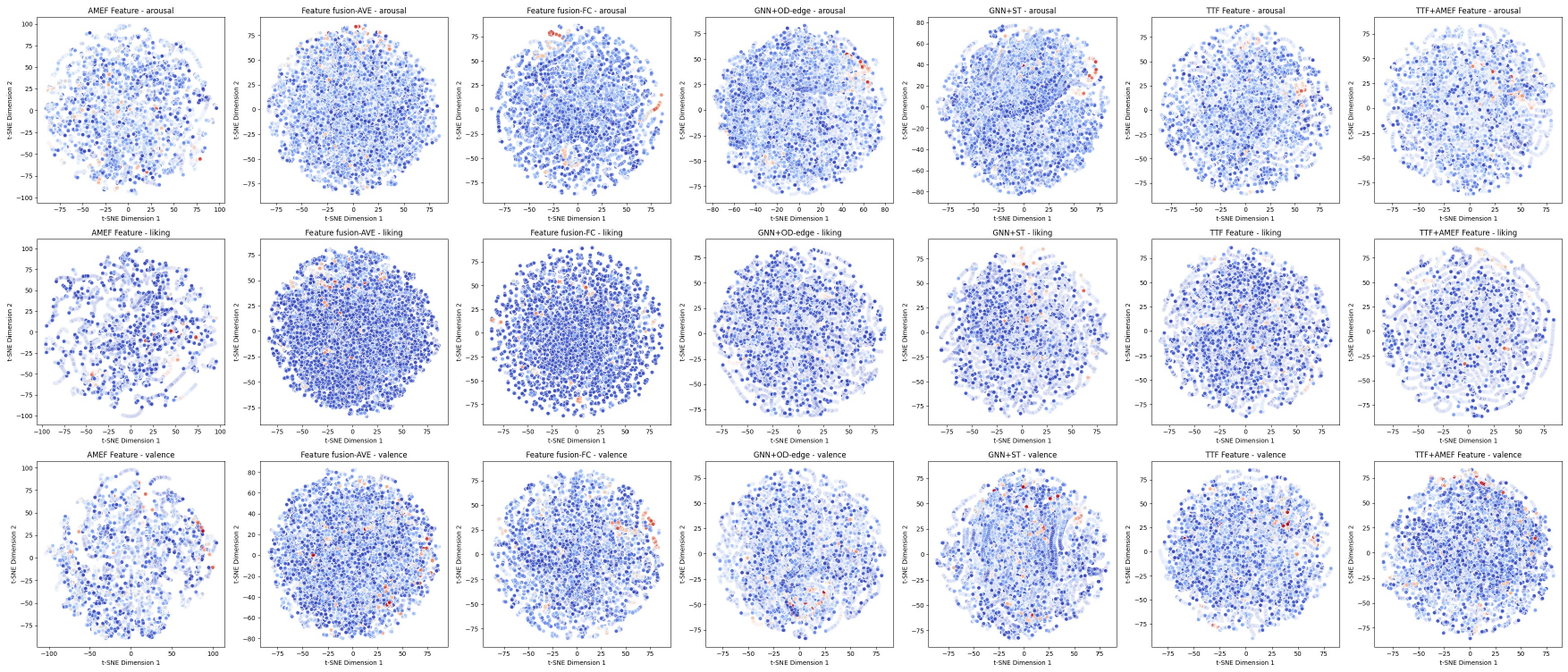}}
	\vspace*{8pt}
	\caption{Ablation experiments on the DE+HU cultural dataset with t-SNE.}
\end{figure}

\begin{table}[!h]
	\tbl{Different combinations of audio features  results evaluated with CCC Score.}
	{\begin{tabular}{@{}ccccccc@{}} \toprule
			
			\multirow{2}{*}{Audio features}	 & 	\multicolumn{2}{c}{Ger. + Hun.}	 & \multicolumn{2}{c}{German}	 & \multicolumn{2}{c}{Hungarian} \\ 
			& 	TTF	 & 	AMEF	 & 	TTF	 & 	AMEF	 & 	TTF	 & 	AMEF	 \\ 
			\colrule
			& 	\multicolumn{2}{c}{Arousal}	 & 		\multicolumn{2}{c}{Arousal}	 & 		\multicolumn{2}{c}{Arousal}	 \\ 
			MFCC+eGeMAPs+BoAW\_e+DS	 & 	\textbf{0.370}	 & \textbf{0.366}	 & 	\textbf{0.443}	 & 	\textbf{0.450}	 & 	\textbf{0.372}	 & 	\textbf{0.420}	 \\ 
			MFCC+eGeMAPs+BoAW\_e	 & 	0.283	 & 	0.341	 & 	0.404	 & 	0.399	 & 	0.364	 & 	0.358	 \\ 
			eGeMAPs+BoAW\_e	 & 	0.363	 & 	0.284	 & 	0.370	 & 	0.374	 & 	0.316	 & 	0.305	 \\ 
			& \multicolumn{2}{c}{Valence}	 & 	\multicolumn{2}{c}{Valence}	 & 	\multicolumn{2}{c}{Valence}	 \\ 
			MFCC+eGeMAPs+BoAW\_e+DS	 & 	0.261	 & 	\textbf{0.292}	 & 	\textbf{0.433}	 & 	0.338	 & 	0.198	 & 	\textbf{0.326}	 \\ 
			MFCC+eGeMAPs+BoAW\_e	 & 	0.269	 & 	0.291	 & 	0.414	 & 	0.350	 & 	0.339	 & 	0.272	 \\ 
			eGeMAPs+BoAW\_e	 & 	\textbf{0.328}	 & 	0.252	 & 	0.352	 & 	\textbf{0.397}	 & \textbf{	0.352}	 & 	0.178	 \\ 
			& 	\multicolumn{2}{c}{Linking}	 & \multicolumn{2}{c}{Linking}		 & 	\multicolumn{2}{c}{Linking}	 \\ 
			MFCC+eGeMAPs+BoAW\_e+DS	 & 	\textbf{0.174}	 & 	0.120	 & 	\textbf{0.183}	 & 	0.115	 & 	\textbf{0.182}	 & 	\textbf{0.199}	 \\ 
			MFCC+eGeMAPs+BoAW\_e	 & 	0.134	 & 	\textbf{0.121}	 & 	0.161	 & 	0.097	 & 	0.151	 & 	0.149	 \\ 
			eGeMAPs+BoAW\_e	 & 	0.151	 & 	0.15	 & 	0.153	 & 	\textbf{0.161}	 & 	0.129	 & 	0.155	 \\ 
			\botrule
	\end{tabular}}
\end{table}

In summary, we found that the TTF feature and the AMEF feature achieved good results in the results, while the TTF feature was superior to the AMEF feature. In order to test our research hypothesis that TTF and AMEF can better capture the best relational cues between different types of speech features, we conducted an experiment where TTF features and AMEF features were fused with two, three, and four sets of speech features, respectively.In this fusion, we discarded the speech features that yielded the worst results: three (MFCC, BoWA-M, DS), two (BoWA-M, DS), and one (BoWA-M).By examining the CCC scores of the prediction outcomes, we observed that the scores decreased as the number of retained feature types diminished.This suggests that even when the prediction performance of a single type of speech feature is below average, our method can still identify relational clues among different types of speech features that are beneficial for enhancing prediction performance.
Consequently, our approach can further improve prediction performance.

\section{Conclusion}

This paper presents a novel approach to speech feature fusion that employs graphical representation learning and graphical neural networks to model speech features derived from disparate datasets.The objective of this study is to investigate the impact of various speech features on speech emotion recognition in the SEWA dataset. Initially, features are extracted from each frame of the input audio in order to obtain five features that capture different aspects of speech emotion information. Subsequently, these features are processed through a two-layer long short-term memory (LSTM) network to extract vectors. This paper presents an innovative approach to speech feature fusion that uses the multidimensional edge feature approach. This approach combines speech feature vectors into a unified representation that captures the interactions between features through edge information and the uniqueness of features through point information. Furthermore, ablation studies were conducted using different feature merging and fusion methods. The results of the experiments conducted on the SEWA dataset demonstrate the accuracy of the method. Furthermore, two sets of experiments, t-SNE and different combinations, were performed to illustrate the effectiveness of the method. 

The main limitations of this study are: (1) the dataset utilised is limited to the SEWA dataset; and (2) the data features of other modalities, such as video, have not been explored. In the future, we intend to incorporate a wider range of speech features and cross-corpus datasets to further optimise our method.

\section*{Acknowledgments}

(Portions of) the research in this paper uses the SEWA Database collected in the scope of SEWA project financially supported by the European Union's Horizon 2020 Research and Innovation Programmer under Grant agreement No. 644869 (SEWA)

This work was supported by Science and Technology Commission of Shanghai Municipality (22692108300) and the National Natural Science Foundation of China (71672128).

\vspace*{-0.01in}
%\vspace*{-0.3in}
\noindent
\rule{12.6cm}{.1mm}

\end{document}